\documentclass[AMS,STIX1COL]{WileyNJD-v2}

\articletype{Research Article}%

\usepackage{tikz}
\tikzset{
	MyPersp/.style={scale=1.8,x={(-0.8cm,-0.4cm)},y={(0.8cm,-0.4cm)},
    z={(0cm,1cm)}},
	MyPoints/.style={fill=white,draw=black,thick}
		}

\usepackage[algo2e, algoruled, linesnumbered]{algorithm2e}
\usepackage{natbib}
\usepackage{graphicx}
\usepackage{mathtools}
\graphicspath{{Images/}}

\begin{document}

\title{Methods to Quantify Dislocation Behavior with Dark-field X-ray Microscopy Timescans of Single-Crystal Aluminum}

\author[1]{Arnulfo Gonzalez*}

\author[1]{Marylesa Howard}

\author[1]{Sean Breckling}

\author[2]{Leora E. Dresselhaus-Marais}


\address[1]{\orgdiv{Signal Processing and Applied Mathematics}, \orgname{Nevada National Security Site}}
\address[2]{\orgdiv{Physical and Life Sciences}, \orgname{Lawrence Livermore National Lab}}

\corres{*Arnulfo Gonzalez. \email{gonzala@nv.doe.gov}}

\abstract[Summary]{Crystal defects play a large role in how materials respond to their surroundings, yet there are many uncertainties in how extended defects form, move, and interact deep beneath a material's surface. A newly developed imaging diagnostic, dark-field X-ray microscopy (DFXM) can now visualize the behavior of line defects, known as dislocations, in materials under varying conditions. DFXM images visualize dislocations by imaging the very subtle long-range distortions in the material's crystal lattice, which produce a characteristic adjoined pair of bright and dark regions. Full analysis of how these dislocations evolve can be used to refine material models, however, it requires quantitative characterization of the statistics of their shape, position and motion. In this paper, we present a semi-automated approach to effectively isolate, track, and quantify the behavior of dislocations as composite objects. This analysis drives the statistical characterization of the defects, to include dislocation velocity and orientation in the crystal, for example, and is demonstrated on DFXM images measuring the evolution of defects at 98$\%$ of the melting temperature for single-crystal aluminum, collected at the European Synchrotron Radiation Facility. }

\keywords{Image Processing, Feature Tracking, Dislocations, Material Science}

\maketitle

\section{Introduction}

Dark-field X-ray microscopy (DFXM) is a new imaging technique that was developed over the last decade to image specific populations of distortions in materials' crystal lattices \cite{Simons2015}. While a hypothetical ``perfect'' crystal would produce a DFXM image showing only a single flat field of intensity across the image, DFXM images of real materials reveal bright and dark objects that originate from imperfections (defects) in the crystal's lattice. For a sufficiently large crystal grain, a single unprocessed DFXM image shows individual defects along a given crystal plane (for crystals with a low density of defects). Classical DFXM studies have measured static materials by collecting a stack of images that resolve different distortion-fields in a crystal, then reconstructing the full distortion field with post-processing; this has already demonstrated DFXM's ability to resolve important sub-surface defect information that is required to refine material models \cite{Jakobsen2019,Simons2018}. Recent work has begun to use raw DFXM images to time-resolve the evolution of defects deep beneath a material's surface\cite{Dresselhaus2020}. This work has already demonstrated the importance of quantifying the statistics of the defect-related features resolved by DFXM, which requires specialized methods. 

\begin{figure}
    \centering
    \includegraphics[scale=0.5]{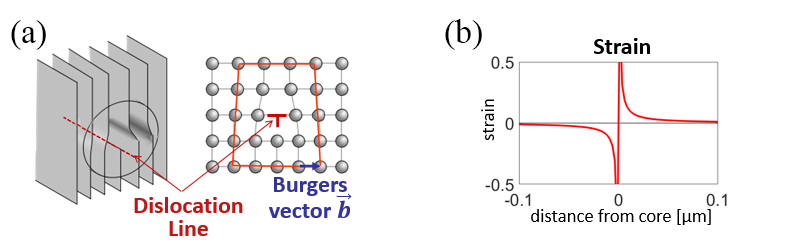}
    \caption{Schematic of a single edge dislocation showing (a) the configuration of the extra truncated plane of atoms that define a dislocation's core, and (b) a plot showing the long-range strain field that emanates from the dislocation core. We note that the full displacement gradient field that defines DFXM's resolution includes contributions from both the strain and rotation tensors, as described in full in \cite{Poulsen2020}.}
    \label{fig:dislocation_schematic}
\end{figure}

A dislocation is a defect of particular importance to materials science, which is shown schematically in Figure \ref{fig:dislocation_schematic} a (left). A dislocation is defined by its core, which is the 2D line that truncates a single extra plane of atoms in the 3D lattice. Long-range displacements fields (comprised of strain and rotation) emanate from the dislocation core, spanning hundreds of nanometers to micrometers in some cases, and DFXM images resolve specific components of that those subtle lattice distortions \cite{Poulsen2020}. As a result, each dislocation appears as a single asymmetric object corresponding to a joined region of dark and light pixels, acting as a bandpass filter of the displacement gradient field, following a similar relation to the strain map depicted in Figure \ref{fig:dislocation_schematic} (right).

\begin{figure}
\centering
  (a)\includegraphics[width = .45\textwidth]{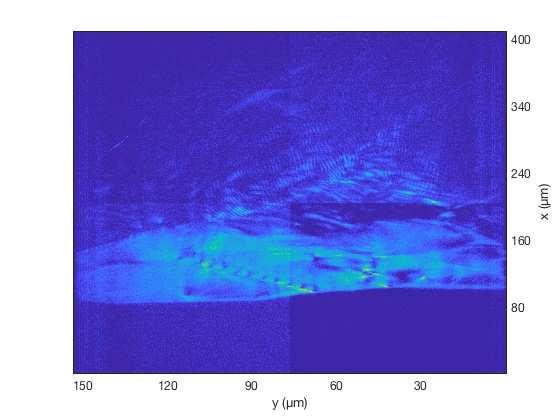}
  (b)\includegraphics[width = .45\textwidth]{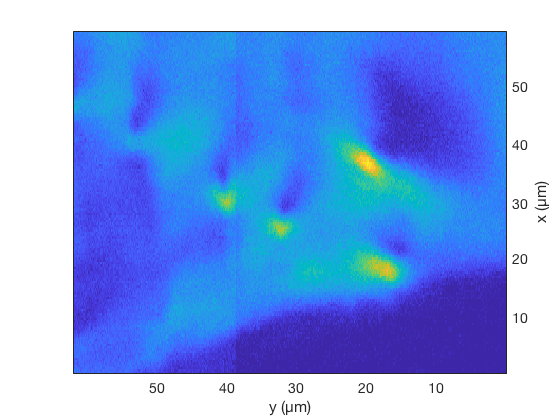}
\caption{The first frame of a timescan experiment, showing full set of dislocations in single-crystal aluminum (a) and zoomed in to a region of interest about active dislocations (b).}
\label{fig:OriginalData}
\end{figure}

Recent experiments described in \cite{Dresselhaus2020} measured time-resolved scans of DFXM at the European Synchrotron Radiation Facility to study how populations of individual dislocations change as a function of time in high-temperature single-crystal aluminum. Each experiment collected timescans at a specific (constant) temperature with increments of 500 frames per scan and a spacing of $\Delta t \approx 0.25$ seconds between frames (precise $\Delta t$ calculated from the timestamps that were recorded for each frame). The first raw image in the timescan is shown in Figure \ref{fig:OriginalData} (left), with and a zoomed-in region about a few dislocations of interest (right). The dislocations are seen as the joined bright/dark pairs and are the primary focus for the analysis workflow in this paper. While it might be conceivable to manually analyze a small set of dislocations from a single timescan movie, these experiments typically perform hundreds of runs and measure thousands to millions of DFXM images. An automated approach to statistically characterize defect behavior is, therefore, essential for rapid advancement in material science.

Even with conventional imaging techniques, quantitative characterization of materials using image analysis is often challenging due to inherent limitations of experiments, for example, acquisition modalities and image-to-image variability from instrumental noise. Statistical computations are further limited by the inability to perfectly replicate irreversible dynamic experiments due to stochastic parameters in the physics. Consequently, many specialized approaches have been developed to analyze imaging data in material science, including multi-scale feature extraction, segmentation, texture analysis, and machine learning \cite{ Jian2010, Fonseca2009, Kumar2015, Bunge1982, DeCost2015, Wei2019}. While processing methods are useful to extract key image features and objects of interest (OOI), to a temporal sequence of images, motion estimation methods are often necessary to precisely track OOIs and quantify their behavior over time \cite{Simmons2019}. In DFXM data, dislocations are comprised of a single bright and dark region pair, the mathematical identification of which is complicated by a fluctuating background intensity profile. To interpret the physics captured by the motion and interactions of individual dislocations resolved in DFXM images, scientists require automated and robust analysis methods that can track the dislocation features in time and space. In this work, we develop an analysis approach to characterize the statistics of dislocation motion and interactions as a function of time to enable materials science studies to interpret the physics in \cite{Dresselhaus2020} and in future similar studies. 

The novelty of this paper is in the effective combination of image processing and computer vision techniques to achieve semi-autonomous object location and tracking within a time-sequence of images, capturing a direct statistical view of defects in the highly-variable DFXM images. The complementary discussion of the quantitative physics and its interpretation are presented in \cite{Dresselhaus2020}. We begin by extracting the five dislocation objects in Section \ref{sec2} using a stationary wavelet transform (SWT) and binarization to identify the bright regions of each OOI, followed by a fast marching method (FMM) to segment the corresponding dark regions. The positions of dislocations are tracked in time with a Kalman filter approach and labeled with a Munkres assignment in Section \ref{sec:DislocTrack}. Each dislocation's motion and behavior are quantified at each time using four quantities of interest (QOIs): position, velocity, acceleration, and orientation in Section \ref{sec:StatChar}. We define  each of the dislocation object positions with centroids, whose  $(x,y)$ positions are rotated approximately 45$^{\circ}$ to orient them along the directions we define as ``glide" and ``climb", respectively (corresponding to the physical mechanism of their motion in each direction \cite{Hirth1992}). The conclusion follows in Section \ref{sec:conclusion}. The analysis is completed using the Image Processing toolbox in Matlab 2018b \cite{Matlab2018}.  

\section{Feature Extraction}\label{sec2}
In this analysis, we manually restrict the full-frame images to capture a specific dislocation interaction event that is of interest to the material's physics (in this case, a lone dislocation inserts into an existing line of dislocations). The selected region of interest (ROI) corresponds to an area approximately 60 x 60 microns in size, given by 410 $\times$ 146  pixels. We analyze only the first 60 frames of the set of 500, as these frames correspond to the most active motion of the OOIs; for the subsequent 440 frames in this ROI, the motion of the OOIs are relatively static. Over our selected  ROI, we resolve the 5 OOIs essential to the physics analysis as they rapidly change in size and direction of motion, and merge and diverge over time.  

\subsection{2D Stationary Wavelet Transform}

To locate the bright regions corresponding to each dislocation OOI in the images, we apply a discrete 2D SWT. The SWT executes a multilevel image decomposition by applying a series of convolutions with low-pass and high-pass decomposition filters, which are associated with a designated orthogonal or biorthogonal wavelet, to the original image array \cite{Mall1989}. Specifically, we use a Daubechies-4 orthogonal wavelet. In this scheme, the original image array is set to an initial approximation coefficients array at a level $j$, which is subsequently used to produce the approximation at next level ($j+1$), as well as the detail coefficient arrays in three spatial orientations: horizontal, vertical, and diagonal. The coefficient arrays, also known as subbands, have dimensions $m \times n \times N$, where $m$, $n$ are the original image array dimensions, and $N$ is equal to the number of levels. As an iterative process, each level produces coarser scales of the image frames. In this case, we applied a 3-level SWT, giving us a wavelet representation of our image consisting of four $2048 \times 2048 \times 3 $  pixel arrays (i.e. the frame dimensions are preserved).  

The SWT is commonly used for noise removal and/or feature extraction, where image arrays are represented by either individual or combined coefficient arrays \cite{Alwan2014,Jumah2013,Zhang2010}. For each frame of the image sequence, we specifically used the horizontal detail coefficients array of the 3rd level to substitute for our original image, as these arrays capture the bright regions of the dislocations with significantly greater intensity relative to the background and omit the noise captured by the remaining detail coefficients. For the first frame in the image series, the third level of the horizontal detail coefficients array is given in Figure \ref{fig:comp}(a).


\subsection{Binarization of SWT on ROI Frames}

\begin{figure}[!htbp]
(a)\includegraphics[width=0.49\textwidth]{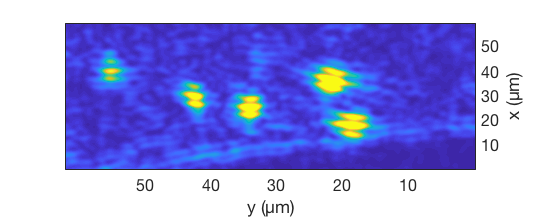}
(b)\includegraphics[width=0.49\textwidth]{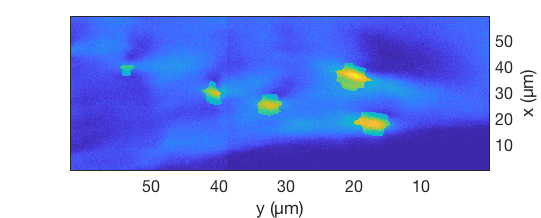}
\caption{The 3rd level of the horizontal detail coefficients array in the stationary wavelet transform corresponding to the ROI of the first frame in the sequence (a), and the extracted OOIs resulting from binarization of (a) with morphological closing plotted on the original frame (b). 
}
\label{fig:comp}
\end{figure}

In order to isolate the bright regions of OOI in the wavelet transformed frames (Figure~\ref{fig:comp}(a)), we convert the SWT frames to binary images. For each of the SWT-ROI frames, connected objects equal to or less than 20 pixels are removed to further reduce image noise. A simple thresholding scheme was applied to binarize the frames. The array median was first subtracted from each of the frames, then the threshold value was set to 3.5 times the standard deviation of the frame (coefficients) array. The thresholding operates on individual wavelet coefficients, setting them to zero when falling below the threshold value:

\begin{equation} 
B_{C} =  {\bf{C}} -  \tilde{C} > 3.5 * \sigma_{C},
\end{equation} 

where $B_{c} $ is the binary array, $C$ is the individual coefficients array, $\tilde{C}$ represents the array median, and $\sigma_{c}$ is the standard deviation of the array. The threshold scaling value was determined empirically to be the minimum value that allows us to capture the OOIs. Following the thresholding, we applied a morphological closing operation to fill in gaps that were found in the remaining connected objects, using a disk-shaped structuring element with a radius of 5-pixels.  The radius chosen is sufficiently small to preserve the original size and shape of objects present in the frames. The resulting extracted OOIs from Frame 1 in Figure~\ref{fig:comp} are overlaid with the corresponding raw image.

\subsection{Segmenting Dark Regions of Dislocations} 

As displayed in Figure~\ref{fig:comp}(a), the bright regions of each of the five dislocation can be effectively extracted using a 2D SWT. To capture the dark regions corresponding to each OOI, each frame of the timescan is segmented using the FMM. The FMM is an efficient numerical method to track the evolution of contours and can be adapted to segmentation by using image features (e.g. grayscale intensity difference) to solve the Eikonal equation \cite{Sethian2013,Forcadel}. In particular, we use Matlab’s \texttt{imsegfmm} function \cite{Matlab2018}, which requires a weight array ${\bf{W}}$, a set of seed points, and a set of threshold values to obtain the segmented OOI from the estimated geodesic distance array. 

The weight array ${\bf{W}}$ consists of weights $w_{ij}$ for each of the pixels in the original ROI image. The value of each weight is inversely proportional to the intensity difference, $d_{ij}$, between each of the image's pixels and the average intensity value of a set of specified seed points (our reference value). That is,
\begin{equation}
w_{ij} = \frac{1}{d_{ij}^{ 2}}, 
\end{equation}
where $i = 410$, $j = 146$  (the size of the image ROI) and the maximum value of $w_{ij}$ is 1.  In this scheme, relatively small weight values indicate background, while larger values indicate foreground. 

The seed points for each frame of the timescan are determined as follows: 

\begin{enumerate}
\item   Extract bright regions of dislocations using SWT, and calculate the centroids \cite{Kaiser1993} .

\item    Starting at each of the five centroids, search pixels in the immediate neighborhood (25 x 25) of the bright regions and record their intensity value. 

\item   For each neighborhood, find the minimum intensity value and set the corresponding pixel as the new seed point. 
\end{enumerate}

The south-easternmost OOI (see the right image in Figure~\ref{fig:comp}) neighbors a dark boundary where the intensity values may be smaller than those found in the dislocation's dark region. To avoid interference from the dark region at the bottom of the image for that particular case, we restrict the search in Step 2 to only the upper right quadrant, relative to the centroid positions for each of the five bright regions. It should be noted that in future implementation of this method to other classes of OOIs, the search region and size may require modification.

The parameter required to threshold the geodesic distance array returned by the FMM is then input as a vector of values tuned to minimally capture each the five OOI. The threshold values are: 0.001, 0.005, 0.01, 0.03 and 0.02, where the smallest (0.01) and largest (0.02) threshold values correspond to an intensity difference of approximately 32 and 7 on the RGB scale, respectively. 

\begin{figure}[!htbp]
(a)\includegraphics[width=.49\textwidth]{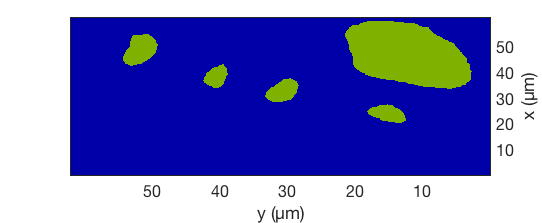}
 (b)\includegraphics[width=.49\textwidth]{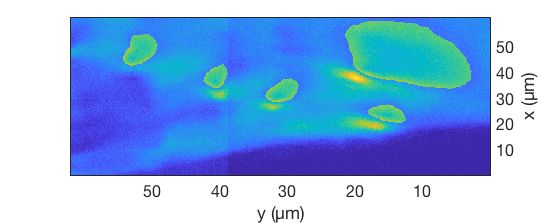}

\caption{The dark regions for each dislocation object, as extracted from Frame 1 using the FMM (a), and an overlay of the dark region OOIs on the original ROI image. 
}
\label{fig:t9}
\end{figure}

As was performed for bright region OOIs, a morphological closing operation was applied using the same disk structuring element with a radius of 5 pixels. In this case, the morphological closing needed to be applied to each OOI individually to avoid erroneous merging. That is, each of the five OOIs were individually mapped to an array of zeros (with the same dimensions as the ROI) and closed; the union of the five arrays is the resulting extraction of the dark region. The resulting dark regions are shown in Figure~\ref{fig:t9}(a) for the first frame and are overlaid on the original ROI frame in (b).

\section{Dislocation Tracking} 
\label{sec:DislocTrack}

\subsection{Composite Objects}

\begin{figure}[!htbp]

(a){\includegraphics[width=0.47\textwidth]{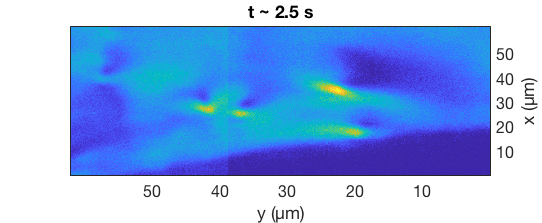}}
(b){\includegraphics[width=0.47\textwidth]{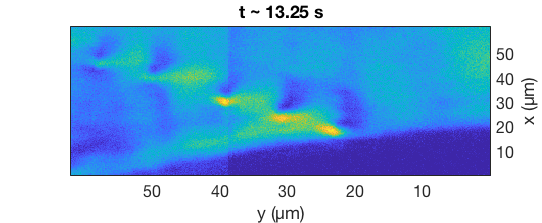}}\\
(c){\includegraphics[width=0.47\textwidth]{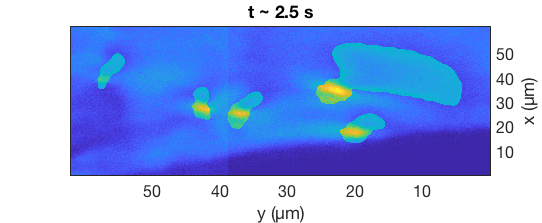}}
(d){\includegraphics[width=0.47\textwidth]{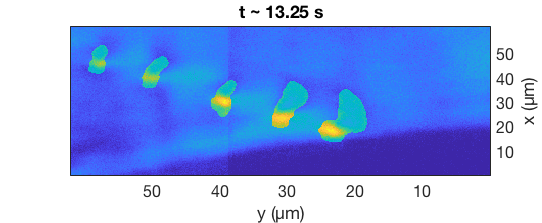}}
\caption{The original image ROI for Frames 10 and 53 are given in figures (a) and (b), respectively. The lone dislocation in Frame 10 has inserted into the dislocation boundary (line) by Frame 53. Using the workflow in Section 1, all five active dislocations are identified and segmented for Frames 10 and 53 in images (c) and (d), respectively. 
}
\label{fig:main}
\end{figure}

After the five bright and dark regions were extracted over the entire sequence of frames, they were combined into the full OOIs by adding the frame arrays. Subsequently, a Gaussian filter, with a standard deviation of 2, was applied to each frame to smooth sharp corners and remove small artifacts introduced by combining the objects \cite{Getruer2013}. The preceding steps produce a set of 60 binarized frames, where the OOIs have been isolated from the background. Using the binarized frames, the OOIs are then tracked according to their centroid positions. In Figure~\ref{fig:main} we show the OOIs overlaid on frames 10 (c) and 53 (d), as well as their corresponding unaltered frames ((a) and (b), respectively). 

\newpage

\subsection{Tracking Dislocation Motion with a Kalman Filter} 

Using the full sequence of binarized frames, the centroid positions of the OOIs are tracked over time by applying a Kalman Filter (KF) to each of the 5 dislocations. Mathematically, the KF is an estimator used to make predictions followed by corrections for states of linear processes, in a manner that minimizes the mean of the square error \cite{Hargrave1989,Welch2006}. For our data, we apply a KF that assumes linear motion of the OOIs between frames, which is a reasonable approximation in this case, as the sampling rate is sufficiently fast compared to the velocity of each dislocation for this assumption to hold. Applying a kinematic model, the KF iteratively predicts the position and velocity of the dislocations in each frame $i$ using the kinematic equations \cite{Weng2016,Sahbani2016}:
\begin{equation}
{r}_{i} = r_{i-1} +  \  v_{i-1} t \ + \frac{a}{2} t^2,
\end{equation} 
\begin{equation}
{v}_{i} = v_{i-1} +  \  at,
\end{equation} 
where acceleration is assumed to be constant, $r = (x,y)$, and $v = (\dot{x},\dot{y})$.
The position and velocity values are incorporated into the KF via the state vector $\bar{X}$ and are predicted using the state dynamic equation:

\begin{equation} 
\bar{X}_{i} = {\bf{A}}  \  X_{i-1} + {\bf{B}} \  u_{i-1} + {\bf{\Eulerconst_{X}}},
\end{equation} 
or equivalently,
 \begin{align}
     \begin{bmatrix}
           x_{i} \\
           y_{i} \\
           \dot{x}_{i} \\
           \dot{y}_{i}         
           \end{bmatrix} 
         &= 
          \begin{bmatrix}
           1 & 0 & dt & 0 \\           
           0 & 1 & 0 & dt \\
           0 & 0 & 1 & 0 \\
           0 & 0 & 0 & 1\\
          \end{bmatrix}  
          \begin{bmatrix}
           x_{i-1} \\
           y_{i-1} \\
           \dot{x}_{i-1} \\
           \dot{y}_{i-1}  \\       
           \end{bmatrix}     
            \begin{bmatrix}
           \frac{dt^{2}}{2} \\
          \frac{dt^{2}}{2}  \\
           dt \\
           dt  \\       
           \end{bmatrix}   
           \cdot a  +
            \begin{bmatrix}
             \frac{dt^{4}}{4} & 0 &   \frac{t^{3}}{2} & 0 \\           
           0 & \frac{dt^{4}}{4} & 0 &   \frac{t^{2}}{2}\\
             \frac{dt^{3}}{2} & 0 &   {dt^{2}} & 0 \\
           0 &  \frac{dt^{3}}{2} & 0 &   \frac{dt^{2}}{2}\\ 
           \end{bmatrix},
  \end{align}
where $\bf{A}$ is the state transition matrix, {\bf{$B$}} is the control input matrix applied to the control vector $u$, and ${\bf{\Eulerconst_{X}}}$ is an error term expressing the variance in the behavior of the system. The time step is set to $dt=1$, and the velocity and acceleration are set to initial values of zero. After applying these conditions, the resulting state measurement equation can be expressed as:

\begin{equation}
\bar{Z}_{i} = {\bf{H}} \bar{X}_{i-1}  + {\bf{\Eulerconst_{Z}}},
\end{equation} 
or equivalently,
\begin{align}
     \begin{bmatrix}
           x_{i} \\
           y_{i} \\        
           \end{bmatrix} 
         &= 
          \begin{bmatrix}
           1 & 0 & 0 & 0 \\           
           0 & 1 & 0 & 0\\
          \end{bmatrix}  
          \begin{bmatrix}
           x_{i-1} \\
           y_{i-1} \\       
           \end{bmatrix}  +     
            \begin{bmatrix}
           \sigma^{2}_{x} & 0 \\
            0 & \sigma^{2}_{y}\\
           \end{bmatrix},
  \end{align}
   
where $\bar{Z}$ is the measurement in frame $i$, and ${\bf{H}}$ is the observation matrix. $\Eulerconst_{Z}$ is the measurement noise covariance matrix, with the standard deviations $\sigma^{2}_{x}$  in the horizontal direction, and $\sigma^{2}_{y}$ in the vertical direction set to 5. For a complete description of the KF algorithm, see \cite{Hargrave1989,Welch2006}.  

\begin{figure}[!htbp]
\centering
  \includegraphics[width=.49\textwidth]{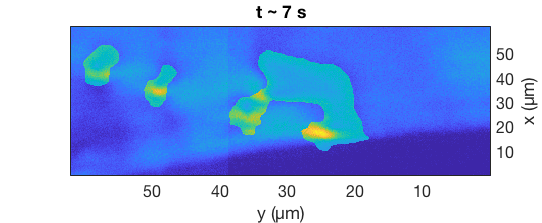}

\caption{Example of a merger, which occludes the bright region of one of the dislocations. In cases like the one shown here, the KF allows us to continue to track the centroids of the bright and dark regions.  
}
\label{fig:t12}
\end{figure}

Ideally, the KF would be applied directly to the dislocations as single composite objects (comprised from the bright and dark regions), however, in this experiment the behavior of dislocations causes them to sometimes merge (events we call ``mergers'') or occlude regions of the the bright and dark regions for in select frames. In these cases, the feature extraction methods applied as above -- specifically using the bright centroids to inform the FMM -- fail because we do not have centroid values for all 5 OOI over the entire 60 frames. Therefore, for frames in which the bright or dark regions cannot be not detected (or are ambiguously detected), we substitute their respective centroid values with KF predictions to complete the object tracking. By using the KF predictions to substitute actual centroid positions, seed locations and unambiguous labels can be maintained for all dislocations. Given the short duration of the image sequence, we define OOIs as objects that can be tracked for 3 observations, and remove their tracks if they are missing from 3 frames.

\subsection{Assignment Algorithm} 

Once positions are identified for each OOI, a Munkres algorithm \cite{Kuhn1955} is used to uniquely label the individual dislocations in each of the 60 frames.  
The initial frame is arbitrarily labeled as dislocations 1-5 based on the centroid positions, $(x,y)$, and the Munkres method labels all subsequent frames by assigning a dislocation based on the  shortest-distance between the detected OOI and the KF-predicted centroid positions. That is, for  a given  frame, we take the centroid position for each dislocation and calculate the Euclidian distance between it and every one of the the centroid positions predicted by the KF for the composite and/or bright and dark regions in the subsequent frame, assigning  the centroid position based on its shortest predicted distance. The algorithm is applied as follows:
 
\begin{equation}        
         D_{x} =  x^{T}_{det.}  - x_{pred.} ,      
\end{equation}

\begin{equation} 
         D_{y} = y^{T}_{det.}  - y_{pred.},
\end{equation}         

\begin{equation}
         {\bf{D}} = \sqrt{D_x.^{2} + D_y.^2},
\end{equation}

\vspace{3mm}

 where $D_{x}.^{2}$ and $D_{y}.^{2}$ indicate element-wise powers, and ${\bf{D}}$ is a $ j \times k $ array (where $j,k =5$ for the five dislocations) representing the Euclidean distances between the detected and predicted dislocation centroids in each frame. The resulting minimum distance is calculated as  

\begin{equation}
min({\bf{D}})   = [ min(d_{1,k}),  min(d_{2,k}),  min(d_{3,k}),  min(d_{4,k}), min(d_{5,k}) ], 
\end{equation}

where the detected dislocations are then ordered according to $argmin({\bf{D}})$.

\section{Characterization of Dislocation Behavior in DFXM Images}
\label{sec:StatChar}
Following the extraction of and tracking of OOIs over the entire sequence of frames, the physical behaviors of interest -- position, velocity, and orientation -- can be characterized. In this case, we describe the position of each dislocation based on its components along the climb and glide directions, corresponding to the different mechanism by which the dislocations must take to move in either direction \cite{Hirth1992}. The relative positions of the dark and bright region for each dislocation is not consistent in each frame, as may be seen by comparing the dislocation objects in Figure~\ref{fig:main}(c) and (d). We define ``orientation'' of each dislocation as the angle between the line connecting the centroids of corresponding bright and dark regions and the horizontal axis, assuming positive angles are counter-clockwise. 

\subsection {Position and Velocity}
\begin{figure}
  (a)\includegraphics[width=.49\textwidth]{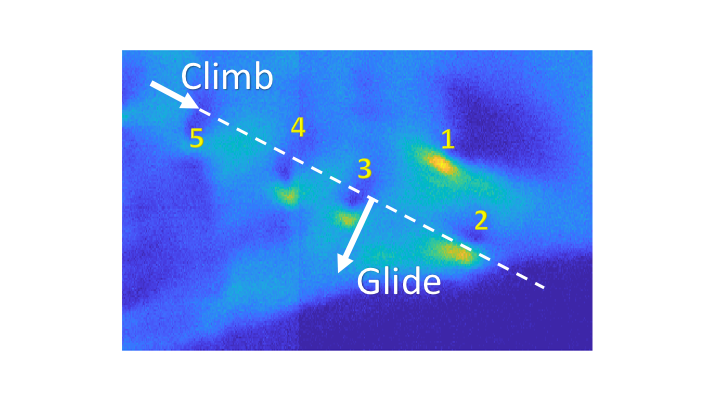}
  (b)\includegraphics[width=.49\textwidth]{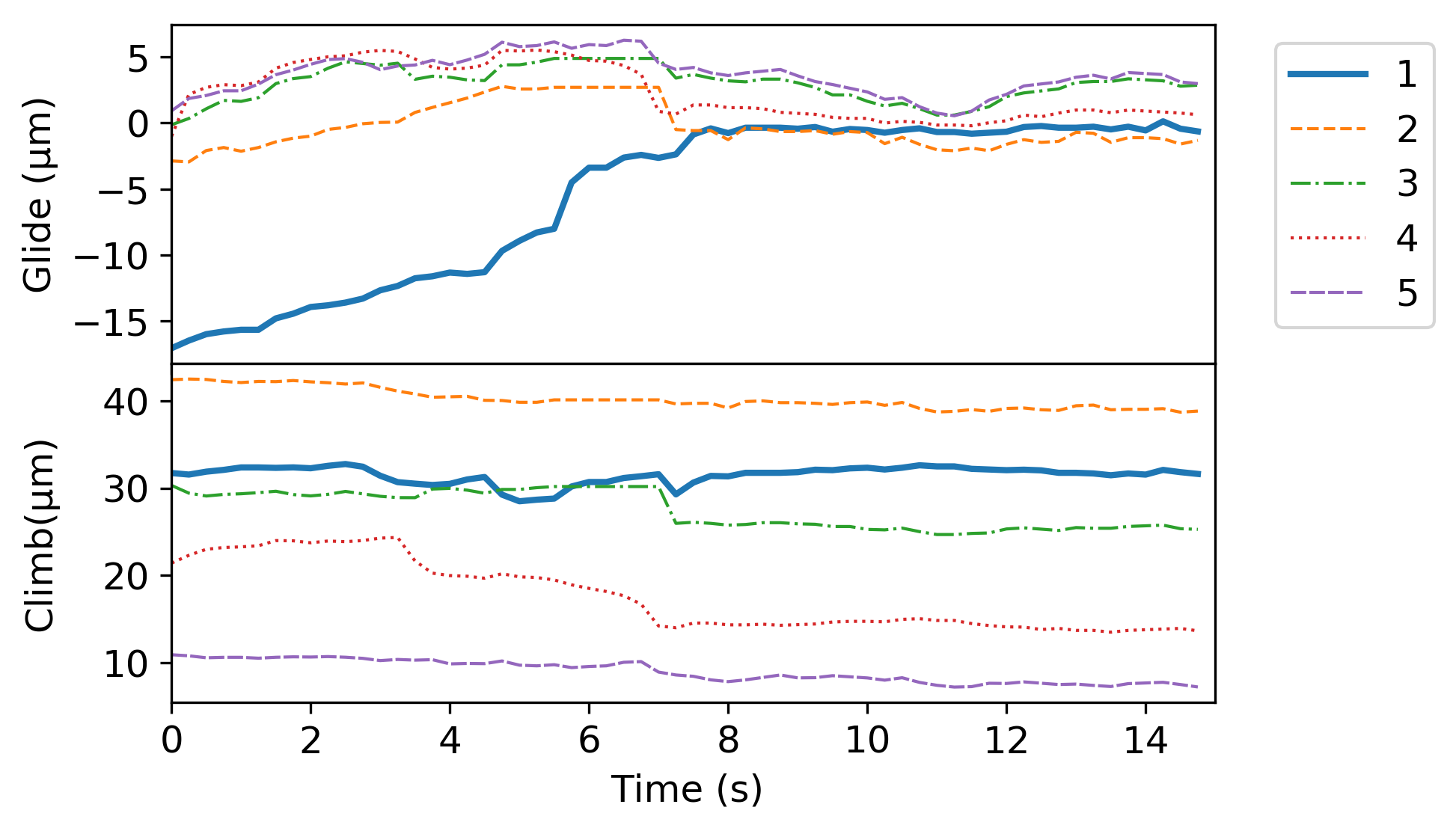}
\caption{Labeled assignment for dislocations and visual representation of the glide and climb reference axes (a). The climb and glide positions over time for the 5 dislocations (b). 
}
\label{fig:climb_glide}
\end{figure}

Each dislocation moves via specific mechanisms that require different energy costs, or activation energies, for specific directions \cite{Hirth1992}. For this reason, physical analysis of the dislocation's motion requires that it be divided into the position and velocity components along two specific directions, defined as the \textit{glide} and \textit{climb} directions, based on the orientation and physics of the crystalline sample. As shown in by the white arrows in Figure~\ref{fig:climb_glide}(a), glide and climb are equivalent to the x- and y-axes rotated by $\approx 45^\circ$ (counter-clockwise), respectively. The five dislocations are labeled 1-5 for tracking purposes. We note that each pixel along the horizontal axis maps to 75-nm in the sample, while each pixel along the vertical axis maps to 205-nm in the sample, however, we specify the climb and glide directions based on their position in the sample.

The positions of dislocations 1-5, decomposed into their components along the glide and climb directions, are plotted as a function of time in Figure~\ref{fig:climb_glide}(b). The positions were decomposed into their components along the glide and climb directions using dot product operators. As Figure \ref{fig:climb_glide} shows, the climb motion deviates only slightly for most of the dislocations over time, with the primary changes occurring as interactions between dislocations 1 and 3 from 4 to 8 seconds. The glide motion, however, increases for each of the dislocations, most notably for dislocation 1, up until the insertion process is completed at frame 30 (t $\sim$ 8.0 s). Following the insertion,the glide motion appears relatively stable. 

\begin{figure}
(a)\includegraphics[width=3.5in,height=2in]{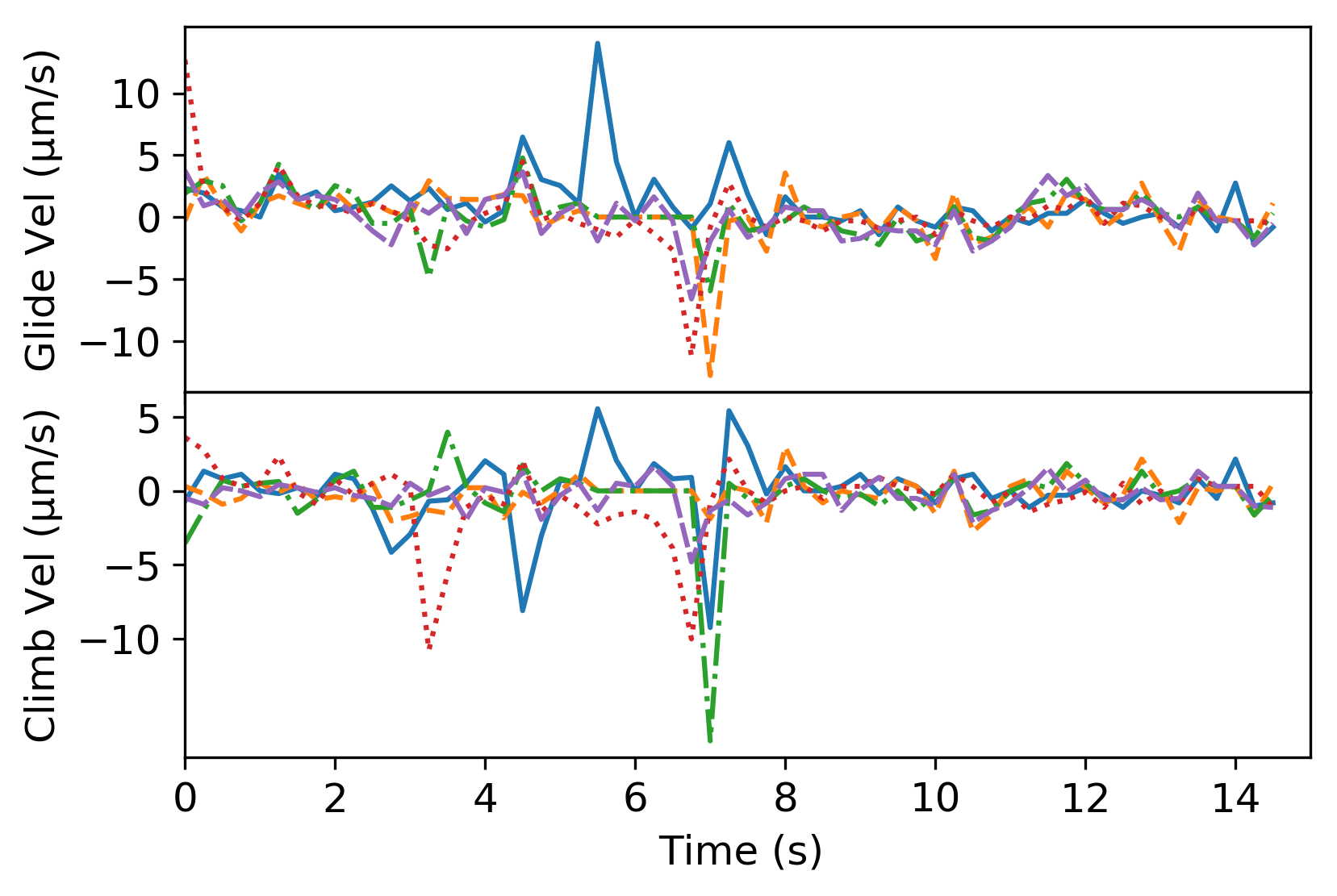}
(b)\includegraphics[width=3.5in,height=2in]{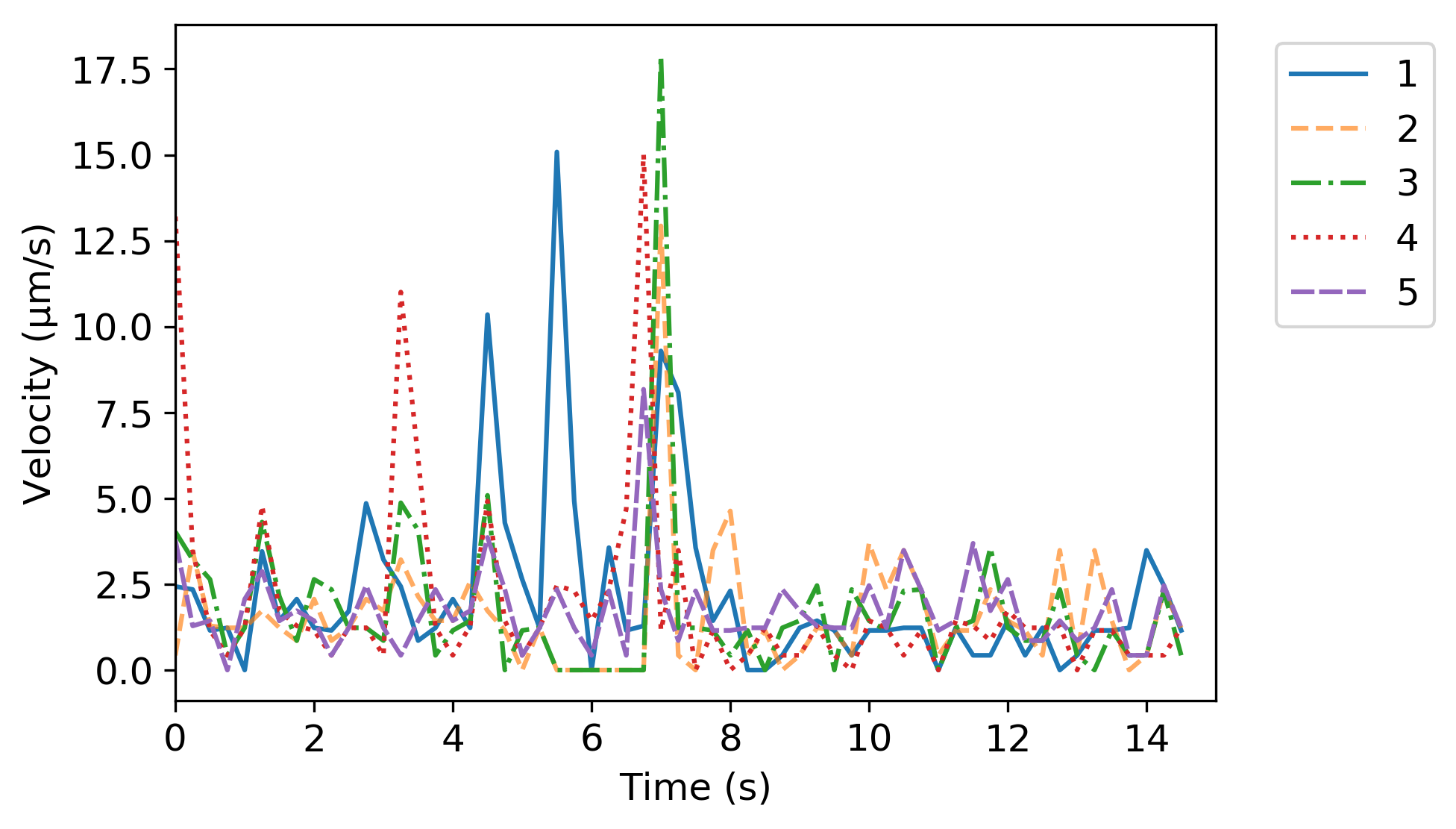}
\caption{Dislocation velocity for the glide and climb directions (a), and absolute velocity (b), plotted as a function of time for the first 60 frames.}
\label{fig:vel}
\end{figure}

The dislocation velocity is calculated by measuring the displacement of centroids, in micrometers, between subsequent frames. Figure~\ref{fig:vel} shows velocity in both glide and climb directions (a), as well as absolute velocity (b). Large changes in the velocity of the dislocations (in both glide and climb directions) appear to be limited to the insertion process, which is completed by $t \sim 8.0 $s. For dislocations 2 and 3, the maximum (negative) velocity occurs at  $t \sim$ 7s, while for dislocations 4 and 5 it occurs in the immediately preceding frame at t $\sim$ 6.75 s, even though it appears that dislocations 4 and 5 are not directly involved in the insertion process. 

\subsection{Orientation}
\begin{figure}
(a)\includegraphics[width=3.5in,height=2.3in]{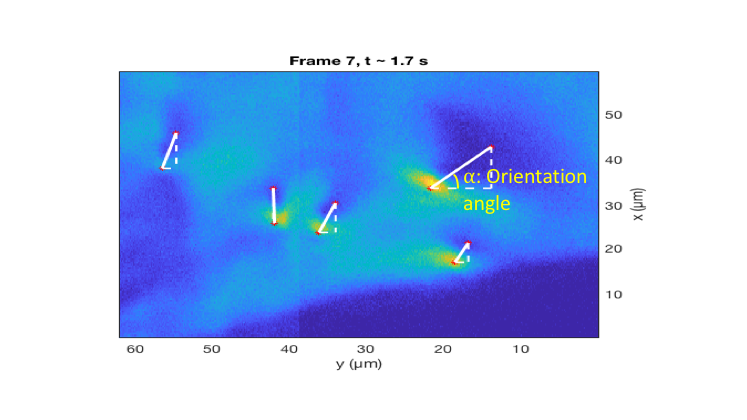}
(b)\includegraphics[width=3.0in,height=2in]{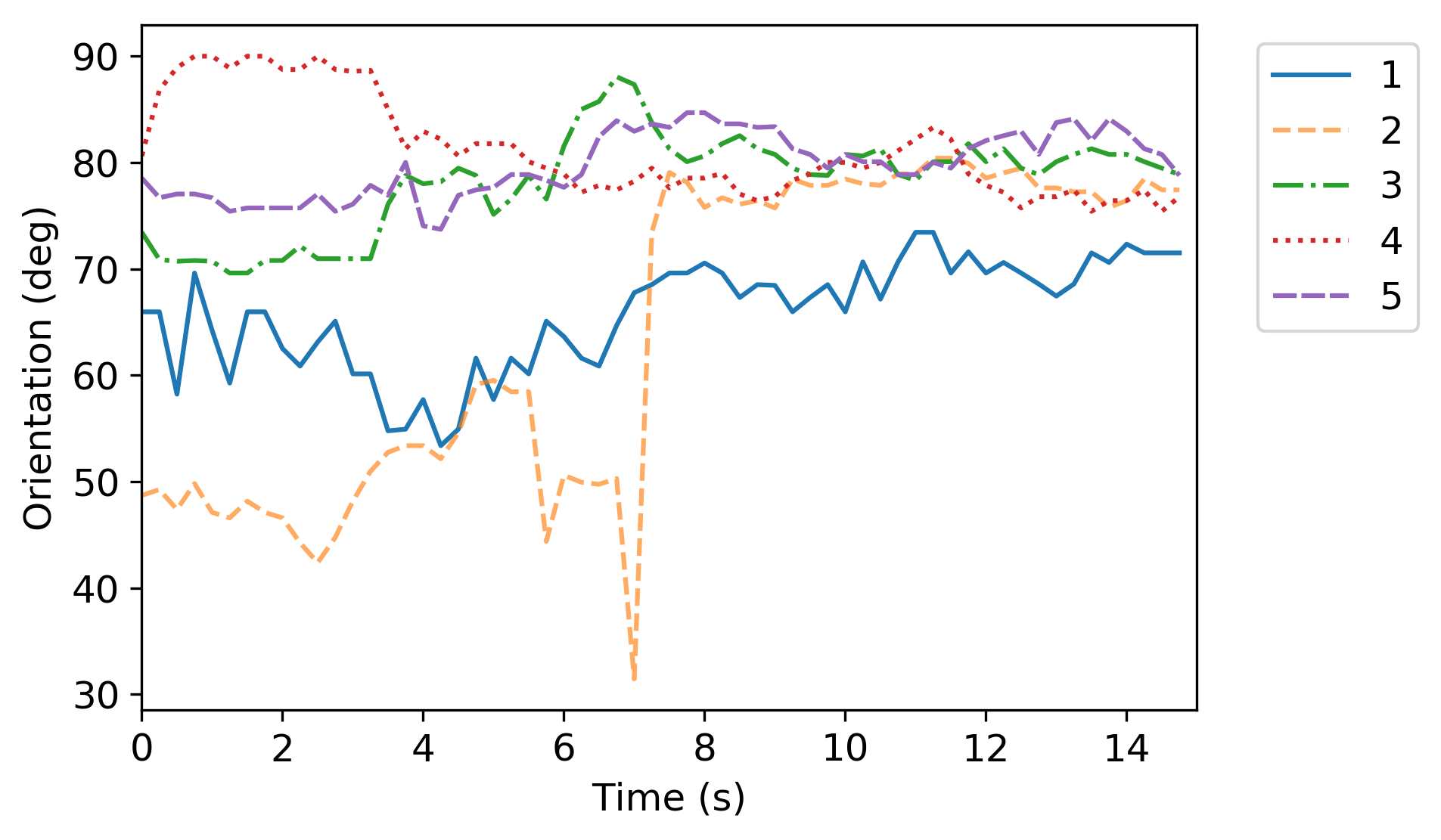}
\caption{Frame 1 is shown with a thick white line connecting the red dots that plot the measured bright and dark centroids and indicating the orientation angle for Dislocation 1 (a). The orientation angles corresponding to all give dislocations are plotted over the entire sequence, providing a figure of merit for the amount of interaction between neighboring dislocations (b).}
\label{fig:orient}
\end{figure}

As we previously defined, the orientation of each dislocation is the angle between the line connecting the bright and dark region centroids and the horizontal axis, in the counter-clockwise direction. Figure~\ref{fig:orient}(a) shows the line and corresponding angle for the dislocation orientation in Frame 1 from the sequence. We track the orientation angle of all five dislocations through time in Figure~\ref{fig:orient}(b), demonstrating the changes to interactions between the distortion fields surrounding each dislocation. When two dislocations are sufficiently close to each other, their surrounding displacement fields can add either constructively or destructively; in our case, the orientation angle demonstrates changes in the interactions between the adjacent dislocations. We observe a relatively sharp contrast between the orientation values for each of the dislocations early in the sequence, particularly when comparing the progressions of dislocations 2 and 4. While the strong dislocation interactions (described fully in \cite{Dresselhaus2020}) cause significant variation in the orientation angles, after dislocation 1 inserts into the array, at approximately $t=8.0$ seconds, the orientations remain fairly stable and much more similar to one another.

\begin{figure}[!htbp]
\centering
(a)\includegraphics[width=.43\textwidth]{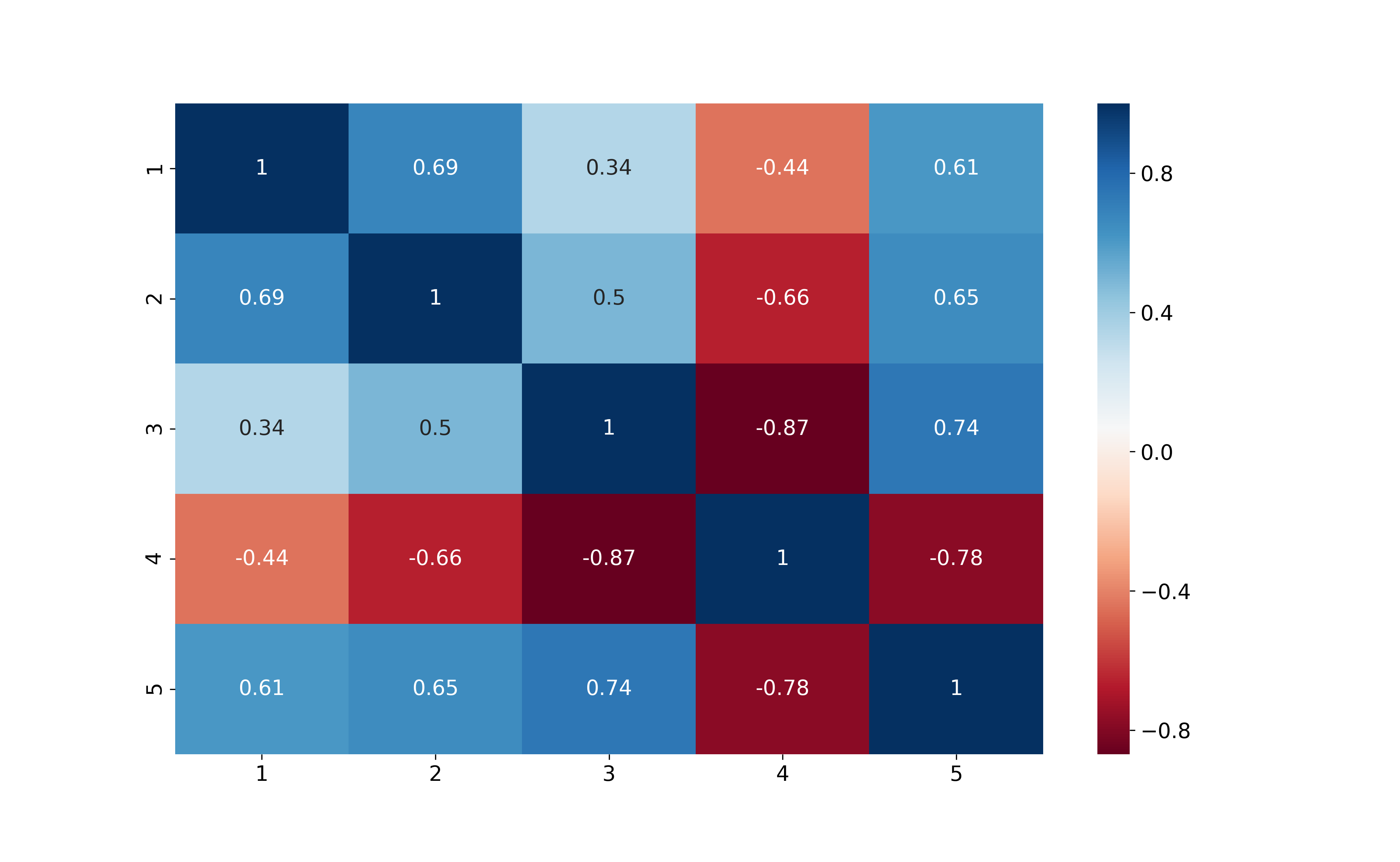}
(b)\includegraphics[width=.48\textwidth]{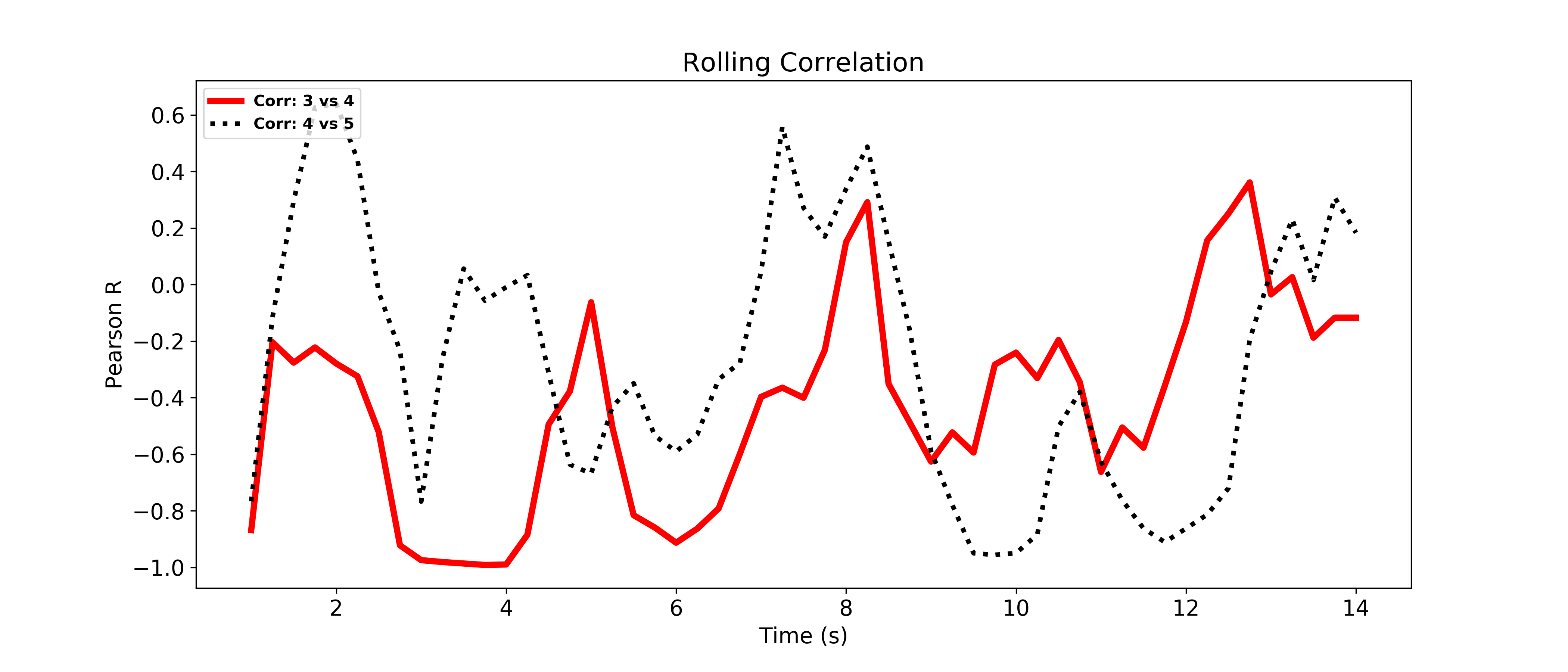}
\caption{A heatmap displaying the global Pearson correlation coefficients for dislocation orientation (a), and the rolling correlation coefficients for orientation measured between dislocations 3 \& 4 and 4 \& 5 (b). 
}
\label{fig:heatmap}
\end{figure}

To resolve correlations between the orientation changes between each pair of dislocations, we show a heatmap of the global Pearson correlation coefficients for dislocation orientation in Figure~\ref{fig:heatmap}(a). This representation shows that there are relatively weak relationships between the orientation of each of the dislocations, with the exception of dislocation 4, which anomalously has a negative correlation with each of the other dislocations. The global correlation coefficient values between dislocations 3 and 4 (-0.87), and 3 and 5 (-0.79), qualify as strong correlations. As such, to further test these relationships, we calculate the rolling coefficient values for the stated pairs and plot them in Figure~\ref{fig:heatmap}(b). Notably, between 3 to 4 seconds, dislocations are almost perfectly negatively correlated -- as the orientation angle of 3 is decreasing from approximately 90 to 80 degrees, angle 4 is increasing from approximately 70 to 80 degrees.

\section{Conclusion}
\label{sec:conclusion}

Dark field X-ray microscopy (DFXM) is a novel imaging diagnostic that allows material scientists to resolve the structural behavior of crystal lattices at the mesoscale. New developments in DFXM have enabled it to visualize the behavior of dislocations over time. The image data requires quantified information about defect behavior to supplement physics models and our understanding of behavior in different materials and environments. The approach presented here demonstrates our ability to identify and locate dislocations in the images and to track them over a sequence of images collected over time. Using the information we capture about each dislocation, we demonstrate progressions of the dislocations in position, velocity, and orientation as the dislocations interact, providing important opportunities to connect DFXM data to the materials science with statistical sampling.

Our approach combines several signal and image processing techniques, and the efficacy of this approach is demonstrated by application to a timescan DFXM video data set of single-crystal aluminum, collected at the European Synchrotron Radiation facility. Beginning with a 2D stationary wavelet transform, we extract the bright regions of each dislocation by representing the original DFXM timescan frames as 3rd level horizontal detail coefficient arrays. Using the centroid locations of the extracted bright regions, we initiate segmentation of the dislocations' dark regions via a seeded fast marching method. This approach allows us to effectively track dislocations both as composite objects and as distinct bright and dark regions. We highlight the importance of tracking these dislocations as split regions, noting that their orientation is defined according to this treatment. Tracking the dislocations allows us to quantify their motion and interactions. By applying a Kalman filter, we track the position of each dislocation, even in cases when they merge or occlude. Following tracking, we can quantify behavior of the defects using position and velocity as a function of the climb and glide planes, as well as orientation versus time. While the time resolution of the data allowed us to successfully apply Kalman filters based on our assumption of nearly linear motion between frames, we note that in future work we expect to incorporate the underlying physics of dislocation motion with more robust non-linear motion estimation models. 

\section{Acknowledgements}

We wish to thank Colin Ophus at Lawrence Berkeley National Laboratory for his insight and helpful discussions about this work. 

This manuscript has been authored in part by Mission Support and Test Services, LLC, under Contract No. DE-NA0003624 with the U.S. Department of Energy and supported by the Site-Directed Research and Development Program, U.S. Department of Energy, National Nuclear Security Administration. The United States Government retains and the publisher, by accepting the article for publication, acknowledges that the United States Government retains a non-exclusive, paid-up, irrevocable, worldwide license to publish or reproduce the published form of this manuscript, or allow others to do so, for United States Government purposes. The U.S. Department of Energy will provide public access to these results of federally sponsored research in accordance with the DOE Public Access Plan (http://energy.gov/downloads/doe-public-access-plan). The views expressed in the article do not necessarily represent the views of the U.S. Department of Energy or the United States Government. DOE/NV/03624-{}-0796. 

Contributions from LEDM were performed under the auspices of the U.S. Department of Energy by Lawrence Livermore National Laboratory under Contract DE-AC52-07NA27344, and the Lawrence Fellowship. 

\bibliography{wileyNJD-AMS}%

\end{document}